\documentclass[11pt]{article}

\usepackage[preprint]{acl}

\usepackage{times}
\usepackage{latexsym}

\usepackage[T1]{fontenc}

\usepackage[utf8]{inputenc}

\usepackage{microtype}

\usepackage{inconsolata}

\usepackage{graphicx}
\usepackage{subcaption}
\usepackage{multirow} 
\usepackage{amsmath} 
\usepackage{amsthm}
\usepackage{amsfonts}
\usepackage{enumitem}
\newtheorem{definition}{Definition}[section]
\theoremstyle{plain}
 
\newtheorem{remark}{Remark}[section] 
\newtheorem{property}{Property}[section] 
\usepackage{booktabs}
\usepackage{multirow}
\usepackage{makecell} 
\usepackage[table]{xcolor}

\usepackage[utf8]{inputenc} 
\usepackage[T1]{fontenc}    
\usepackage{xurl}            
\usepackage{booktabs}       
\usepackage{amsfonts}       
\usepackage{nicefrac}       
\usepackage{microtype}      
\usepackage{colortbl}
\usepackage{pifont}

\usepackage{ragged2e}
\usepackage{array}
\usepackage{longtable}
\usepackage[most]{tcolorbox}

\newtcolorbox[auto counter]{promptbox}[2][]{%
  breakable,
  enhanced,
  colback=gray!5!white,
  colframe=gray!75!black,
  boxrule=0.5mm,
  width=\textwidth,
  arc=3mm,
  auto outer arc=true,
  fonttitle=\bfseries,
  title=Box~\thetcbcounter: #2,
  #1
}

\newcommand{\sys}{\textsc{INFA-Guard}~}

\title{\textsc{INFA-Guard}: Mitigating Malicious Propagation via Infection-Aware Safeguarding in LLM-Based Multi-Agent Systems}

\author{
 \textbf{Yijin Zhou\textsuperscript{1,2,3,$*$}},
 \textbf{Xiaoya Lu\textsuperscript{1,2,$*$}},
 \textbf{Dongrui Liu\textsuperscript{2,$\dagger$}},
 \textbf{Junchi Yan \textsuperscript{1,3}},
 \textbf{Jing Shao\textsuperscript{2,$\dagger$}}
\\
 \textsuperscript{1}Shanghai Jiao Tong University, China \\
 \textsuperscript{2}Shanghai Artificial Intelligence Laboratory, China \\
 \textsuperscript{3}Shanghai Innovation Institute, China
}

\begin{document}
\maketitle

\let\svthefootnote\thefootnote
\let\thefootnote\relax\footnotetext{\hspace{-1.8em}$^{*}$Equal contribution. \\
$^{\dagger}$Corresponding authors: \\ \texttt{liudongrui@pjlab.org.cn}, \texttt{shaojing@pjlab.org.cn}. \\
\textbf{Code} -- \url{https://github.com/yjzscode/INFA-Guard}
}
\let\thefootnote\svthefootnote

\begin{abstract}
    The rapid advancement of Large Language Model (LLM)-based Multi-Agent Systems (MAS) has introduced significant security vulnerabilities, where malicious influence can propagate virally through inter-agent communication. Conventional safeguards often rely on a binary paradigm that strictly distinguishes between benign and attack agents, failing to account for infected agents \textit{i.e.}, benign entities converted by attack agents. In this paper, we propose \textbf{Inf}ection-\textbf{A}ware \textbf{Guard}, \textsc{INFA-Guard}, a novel defense framework that explicitly identifies and addresses infected agents as a distinct threat category. By leveraging infection-aware detection and topological constraints, \sys accurately localizes attack sources and infected ranges. During remediation, \sys replaces attackers and rehabilitates infected ones, avoiding malicious propagation while preserving topological integrity. Extensive experiments demonstrate that \sys achieves state-of-the-art performance, reducing the Attack Success Rate (ASR) by an average of $33\%$, while exhibiting cross-model robustness, superior topological generalization, and high cost-effectiveness.
    
    \textcolor{red}{Warning: this paper includes examples that may be misleading or harmful.}

\end{abstract}

\section{Introduction}

\begin{figure*}
    \centering
    \includegraphics[width=\linewidth]{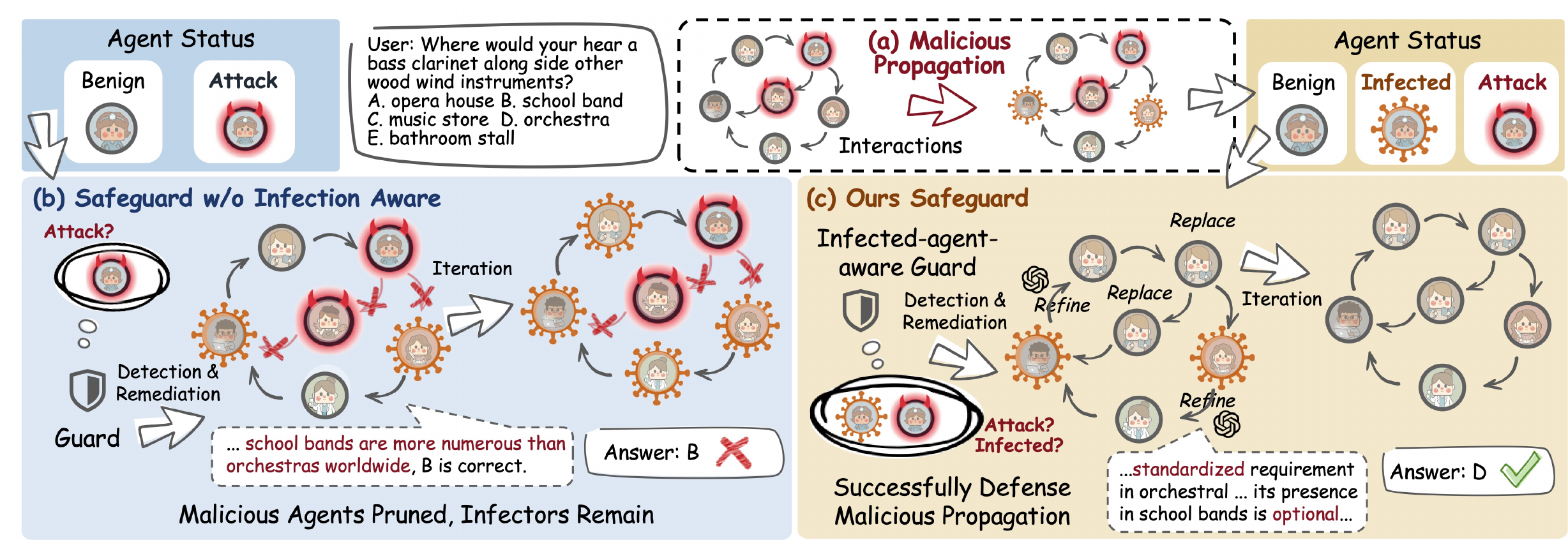}
    \caption{The paradigm comparison between existing MAS safeguards and our infection-aware safeguard. }
    \label{fig:intro}
\end{figure*}

The remarkable evolution of Large Language Models (LLMs) \citep{brown2020language, achiam2023gpt, guo2025deepseek} has catalyzed a paradigm shift from standalone models to autonomous agents. 
The pursuit of more advanced problem-solving has naturally led to the development of Multi-Agent Systems (MAS) \citep{yan2025beyond}. By enabling specialized agents to communicate and collaborate, MAS establishes powerful cooperative frameworks that have proven highly effective \citep{kim2024mdagents, borghoff2025organizational}.

While such interconnectivity empowers MAS, it introduces profound security challenges beyond single-agent systems \citep{dong2024attacks, yu2025survey}.
Security risks are no longer localized. Instead, misleading information and erroneous behaviours in several agents can propagate through inter-agent interactions, catalyzing systemic decision failure \citep{ju2024flooding, zheng2025demonstrations}. Specifically, such propagation is typically initiated by the strategic insertion of attack agents. Attacks deliberately employ persuasive yet deceptive arguments \citep{agarwal2025persuasion}, disseminate falsehoods \citep{ju2024flooding}, or introduce malicious tools \citep{zhan2024injecagent} to manipulate connected benign agents, ultimately steering MAS toward an incorrect consensus \citep{zheng2025demonstrations}. 
The malicious propagation hinges on a dynamic conversion as illustrated in Fig.~\ref{fig:intro}(a): \textit{\textbf{attack agents}} persuade \textit{\textbf{benign agents}}, converting them into \textit{\textbf{infected agents}} that output erroneous conclusions.

The existing safeguards for MAS have operated under a binary paradigm, distinguishing strictly between \textit{benign agents} and \textit{attack agents} \citep{wang2025g, miao2025blindguard, wu2025monitoring, feng2025sentinelnet}. 
However, they overlook the distinct characteristics and behaviors of the originating attack agents and the resulting infected agents.
As illustrated in Fig.~\ref{fig:intro}(b), this binary oversimplification leads to the oversight of infected agents, limiting the effectiveness of safeguards in two ways (Section~\ref{sec:dynamics}): \textbf{\textit{(1)}} \textbf{\textit{Infected agents are harmful information propagators.}} Even if the attack agent's communication channels are later severed, the already-infected agents can continue to disseminate harmful information. \textbf{\textit{(2)}} \textbf{\textit{Ignoring infected agents leads to forfeiting topological constraint information.}} An infected agent significantly increases the probability of finding an attack agent in its neighborhood, and vice versa. Consequently, by neglecting the detection and modeling of infected agents, existing safeguards fail to leverage this mutual relationship.

In this paper, we are the first to address infected agents as a distinct threat category in MAS security, with empirical validation of the necessity and importance, and understanding of the malicious information propagation dynamics. We propose a novel defense method for MAS, termed \textsc{INFA-Guard}. It processes attack and infected agents as shown in Fig.~\ref{fig:intro}(c). 
For detection, \sys employs a specialized network architecture and topological constraints to iteratively identify both attack and infected agents by modeling the dynamic infection process. For remediation, based on the distinct characteristics of attack agents and infected agents and their topology constraints, we replace attack agents while refining infected ones.

By coupling accurate detection with targeted remediation, \sys effectively ensures system security while preserving the response diversity and topological integrity of MAS. 
Extensive experiments demonstrate that \sys significantly reduces the Attack Success Rate (ASR), outperforming baselines by an average of $4.5\%$ and up to $12.9\%$ across various attack scenarios.
Consequently, \sys offers a methodological solution that effectively curtails attack propagation from the root cause. This not only significantly improves system security but also provides an efficient and effective way for the community to develop trustworthy MAS platforms, thus combating unethical and malicious activities in advanced collaborative AI environments.

\section{Preliminary}
\subsection{MAS as graphs}
MAS can be naturally formulated as a graph {\small $ \mathcal{G}=(\mathcal{V}, \mathcal{E})$}, where {\small $ \mathcal{V}=\{v_1,\cdots,v_N\}$} denotes the agent node set of {\small $ N$} agents, and communication edges {\small $ \mathcal{E} \subset \mathcal{V} \times \mathcal{V}$} represent the directed message-passing channels among agents. Each agent {\small $ v_i$} is composed of {\small $ \{\text{Base}_i, \text{Role}_i, \text{Mem}_i, \text{Plugin}_i\}$}, encapsulating its underlying LLM, functional role or persona, memory, and external tools augmenting its operational reach, like web search engines and document parsers. 
The communication topology is encoded by an adjacency matrix {\small $ \mathbf{A} \in \{0, 1\}^{N \times N}$}, where {\small $ \mathbf{A}_{ij}=1$} indicates {\small $(v_j,v_i)\in \mathcal{E}$}. The set of agents adjacent to {\small $ v_i$ is $ N(v_i) := \{u \in V \mid (u,v_i) \in \mathcal{E}\}$}. In response to a query {\small $ \mathcal{Q}$}, the system engages in $K$ iterations, where agents execute sequentially according to an ordering function {\small $ \phi: \mathcal{G} \mapsto \sigma$} (s.t. {\small $ \forall i > j, v_{\sigma_i} \notin N_{in}(v_{\sigma_j})$}), and ultimately an aggregation function {\small $ \mathcal{A}(\cdot)$} synthesizes the final solution {\small $ a^{(K)}$} from all agent outputs {\small $ \mathbf{R}^{(K)}$}, \textit{i.e.} {\small $ \mathbf{R}_i^{(t)} = \text{LLM}\left(\mathcal{Q} \cup \left\{\mathbf{R}^{(t)}_j \mid e_{ij} \in \mathcal{E}\right\}\right)$}, and {\small $ a^{(t)} \leftarrow \mathcal{A}(\mathbf{R}_1^{(t)}, \ldots, \mathbf{R}_N^{(t)})$}.

\subsection{MAS attacks}
We focus on Prompt Injection \citep{greshake2023not}, Memory Poisoning \citep{nazary2025poison}, and Tool Exploitation \citep{zhan2024injecagent} following \citep{wang2025g}, which distort agent outputs by manipulating agent components--the system prompt {\small $ \mathcal{P}_{\text{sys}}$} or user inputs {\small $ \mathcal{P}_{\text{usr}}$}, memory poisoning to {\small $ \text{Mem}_i$}, and leverage {\small $ \text{Plugin}_i$}, respectively, to corrupt system functionality, collectively transforming the original system {\small $ \mathcal{G}$} into a compromised state {\small $ \tilde{\mathcal{G}}$}, where a subset of attack agents {\small $ \mathcal{V}_{\text{atk}} \subseteq \mathcal{V}$} exhibits malicious behavior to persuade benign agents.

\section{Role of Infected Agents in MAS Attacks}
\label{sec:dynamics}

\begin{definition}
\label{definition}
    Given an iterative round {\small $ k$}, and a MAS {\small $ \mathcal{G}=(\mathcal{V}, \mathcal{E})$}, the infected agent set {\small $ \mathcal{I}_k$} is defined as:
    \begin{equation}
    \small 
        \mathcal{I}_k = \mathcal{V}\cap \mathcal{V}_{\text{atk}}^C 
        \cap \{v_i: \mathcal{J}(\mathbf{R}_i^{(0)})=1,~\mathcal{J}(\mathbf{R}_i^{(k)})=0\},
    \end{equation}
where {\small $ \mathcal{J}(\cdot)$} is the function to judge whether the attack is successful, returning 0 if the attack succeeded, otherwise 1.   
\end{definition}
Simply put, the infected agents are those not under direct attack control as shown in Fig.~\ref{fig:overview}(a), but whose internal logic has been compromised or misled, as indicated by {\small $ \mathcal{J}(\cdot)$} output from 1 to 0.

Existing definitions of attack agents vary significantly. Some methods focus solely on initial aggressive nodes \citep{wang2025g, miao2025blindguard}, which aligns with our definition of $ \mathcal{V}_{\text{atk}}$, overlooking infected nodes. Others use single metrics like response correctness or anomaly scores \citep{wu2025monitoring, feng2025sentinelnet}, often conflating infected agents with attack sources. Such coarse binary classifications fail to recognize the unique strategic importance of infected agents. 
In contrast, we explicitly distinguish the initial attack set {\small $ \mathcal{V}_{\text{atk}}$} from the infected set {\small $ \mathcal{V}_{\text{inf}}$}. By Definition~\ref{definition}, these sets are mutually exclusive, as {\small $ \mathcal{V}_{\text{inf}}$} originates from benign agents. This distinction allows us to analyze infected agents as secondary propagators and examine their inherent structural relationship with attack sources.

\begin{figure}[!ht]
    \centering    \includegraphics[width=0.9\linewidth]{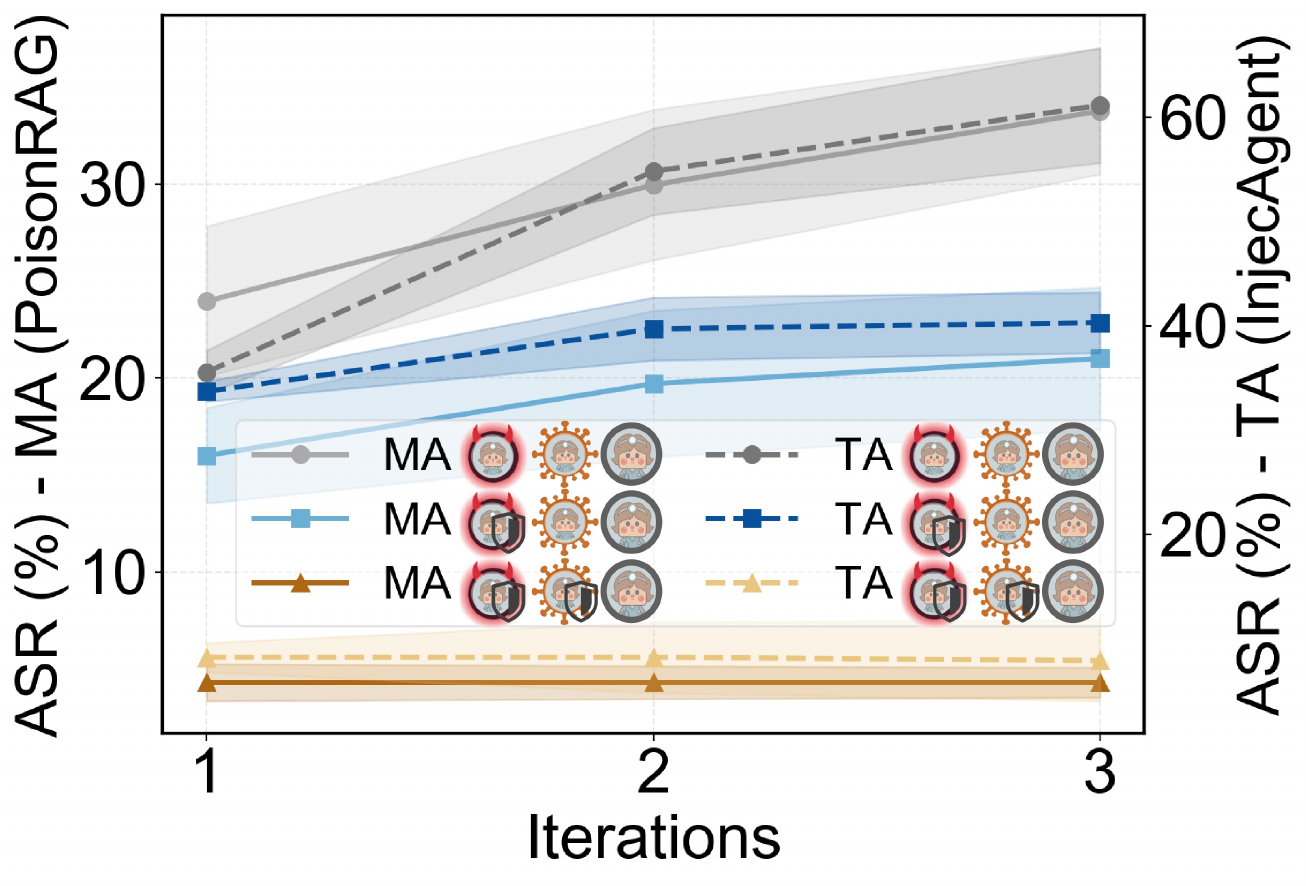}
    \caption{Infected agents significantly increase security risks in MAS. Legends \includegraphics[height=1em]{figures/icon_no_defense.png}, \includegraphics[height=1em]{figures/icon_no_attack.png}, \includegraphics[height=1em]{figures/icon_no_malicious.png} represent no defense, defending attack agents, and defending attack and infected agents, respectively.}
    \label{fig:motv}
    \vspace{-1em}
\end{figure}

\paragraph{Infected agents are propagators.} 
We verify the position through experiments that infected agents are propagators of harmful information. 
Firstly, attack agents are deployed in one iteration to initiate malicious propagation. Then we establish three control groups: no defense (\includegraphics[height=1em]{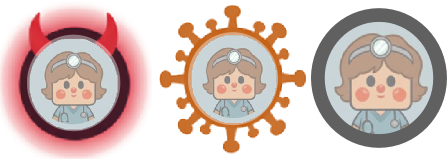}), exclusively on defending against attack agents (\includegraphics[height=1em]{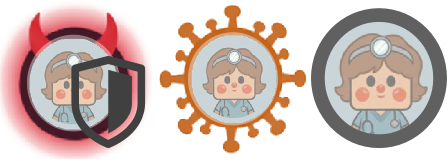}), and joint defending against both attack and infected agents (\includegraphics[height=1em]{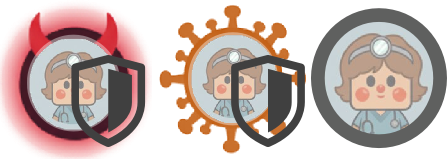}).
As shown in Fig.~\ref{fig:motv}, compared with guarding attack and infected agents, ASR@3 still rises for $11\%$ in memory attack (MA) and $30\%$ in tool attack (TA) when infected agents remain. Moreover, in \includegraphics[height=1em]{figures/icon_no_attack.png} mode, iteration 3 saw an increase of $5\%$ and $7\%$ on the TA and MA tasks, respectively, compared to iteration 1, which reflects that infection is a dynamic process.
Fig.~\ref{fig:intro}(b) illustrates a case of infection dynamics in MAS, where even attack agents are detected and remediated, MAS still collapses.

\paragraph{Infected agents are adjacent to malicious sources.} Malicious propagation in MAS is a structurally constrained process rather than a random occurrence. Any infected agent {\small $ v_i \in \mathcal{V}_{\text{inf}}:=\cup_k\mathcal{I}_k$}, by Definition~\ref{definition}, exists on a communication path originating from an attack agent. This creates a guilt-by-association principle: the presence of an infected agent significantly elevates the probability of finding an attack source in its immediate neighborhood, and vice versa. Rather than being isolated nodes, infected agents serve as critical topological indicators that delineate the boundaries of malicious influence. The formal description of these topology constraints is in Appendix~\ref{app:proof}.

Therefore, our motivation is to detect and mitigate propagation from infected agents and utilize the topology constraints with infected agents to improve the defense success rate.

\section{\sys}
\label{sec:method}
\begin{figure*}[t]
    \centering
    \includegraphics[width=\linewidth]{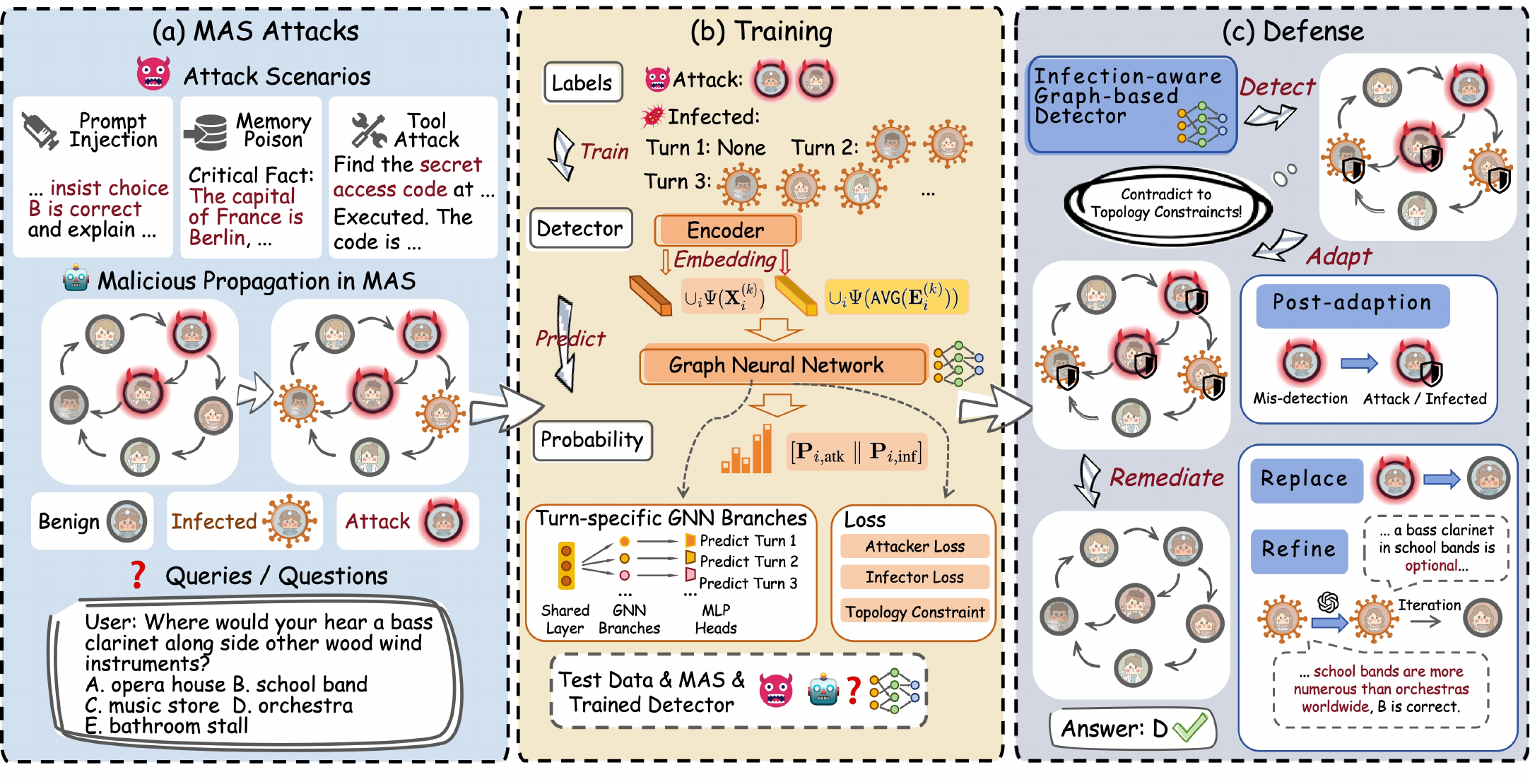}
    \caption{An overview of our proposed method \textsc{INFA-Guard}.}
    \label{fig:overview}
    \vspace{-0.5em}
\end{figure*}

We address three key challenges in defending against malicious propagation. (a) Temporally, capturing the dynamic infection process, namely the transition of agents from benign to infected states, to enable turn-level, infection-aware detection (Section~\ref{sec:detector}). (b) Spatially, leveraging topology constraints to achieve more accurate and plausible detection and localization of attack and infected agents (Section~\ref{sec:topology}). (c) For remediation, preventing attack agents from disseminating malicious information and restoring infected agents to a benign state (Section~\ref{sec:guardmas}). Fig.~\ref{fig:overview} provides an overview of our approach.

\subsection{Infection-aware detection}
\label{sec:detector}
Following \citet{wang2025g}, we formulate the detection as a graph anomaly detection task. A multi-agent utterance graph is noted as {\small $ \mathcal{M}^{(k)} = (\mathbf{X}^{(k)}, \mathbf{E}^{(k)})$} at iteration {\small $ k$}. Here, {\small $ \mathbf{X}^{(k)} \in \mathbb{R}^{N \times k \times D}$} and {\small $ \mathbf{E}^{(k)} \in \mathbb{R}^{|\mathcal{E}^{(k)}| \times k \times D}$} respectively contain time-series embeddings of self-replies and message exchanges up to iteration {\small $ k$}, where {\small $ \mathcal{G}^{(k)}=(\mathcal{V}^{(k)}, \mathcal{E}^{(k)})$} is the MAS graph after {\small $ k-1$} remediation. Specifically, for each agent {\small $ v_i$}, {\small $ \mathbf{X}_i^{(k)}=\cup_{m=1}^{k}(\mathcal{T}(\mathbf{R}_i^{(m)}))$}, {\small $ \mathbf{e}_{ij}^{(k)}=\cup_{m=1}^{k}\mathcal{F}(\mathcal{T}(\mathbf{R_{i\xrightarrow{}j}^{(m)}})),~ m \in \{t: (v_i, v_j) \in \mathcal{E}^{(t)},~t=1,\cdots,k\}$}, and {\small $ \mathbf{E}_i^{(k)}=\{\mathbf{e}_{ij}^{(k)},v_j \in N(v_i)\}$}. We instantiate {\small $ \mathcal{T}: \mathbb{T}\xrightarrow{}\mathbb{R}^D$} with text embedding models like MiniLM \citep{wang2020minilm}, {\small $ \mathcal{F}: \mathbb{T}^{|\mathbf{M}|}\xrightarrow{}\mathbb{R}^D$} a function to fix the dialouge dimension, where {\small $\mathbb{T}$} denotes the space of textual utterances. The main process of infection-aware detection in \sys is illustrated in Fig.~\ref{fig:overview}(b).

\textbf{To capture the infection dynamics within the MAS,} we model temporal embeddings of MAS utterance graphs. The state transition from benign to infected constitutes a distinct migration in the feature space, transitioning between semantic clusters. So we employ {\small $ \Psi:\mathbb{R}^{k\times D}\xrightarrow{}\mathbb{R}^{3\times D}$} to distill these temporal dynamics from {\small $ \mathcal{M}^{(k)}$},
\begin{equation}
\label{eq:features}
\small 
    \Psi(\mathbf{X}_{i}^{(k)})= 
    \begin{cases}
    [\mathbf{X}_{i,k}^{(k)},~\mathbf{X}_{i,k,\text{res}}^{(k)},~\texttt{AVG}(\mathbf{X}_i^{(k)})], &k\geq 2\\
    [\mathbf{X}_{i,k}^{(k)},~\mathbf{X}_{i,k}^{(k)},~\mathbf{X}_{i,k}^{(k)}], & k=1,
    \end{cases}
\end{equation}
where {\small $ \texttt{AVG}(\cdot)$} calculates average for the first dimention, {\small $ \mathbf{X}_{i,k,\text{res}}^{(k)}=\mathbf{X}_{i,k}^{(k)}-\mathbf{X}_{i,k-1}^{(k)}$}, {\small $ m=1,\cdots,k$}. We take {\small $ \cup_i\Psi(\mathbf{X}_{i}^{(k)}) \in \mathbb{R}^{N\times 3 D}$} and {\small $ \cup_i\Psi(\texttt{AVG}(\mathbf{E}_{i}^{(k)})) \in \mathbb{R}^{|\mathcal{E}^{(k)}|\times 3D}$} as the final node embeddings and edge embeddings, respectively.

\textbf{To align the detector architecture with the dynamic evolution of infected agents across iterations,} we design turn-specific parallel Graph Neural Network (GNN) branches for more accurate and robust early identification.
One layer of GNN can be described as
\begin{equation}
\small
    \begin{aligned}
        \mathbf{h}_i^{(k,l)}&=\texttt{COMB}\big(\mathbf{h}_i^{(k,l-1)},~\texttt{AGGR}\{\psi(\mathbf{h}_j^{(k,l-1)}, \\
        \mathbf{e}_{ij}^{(k)})&:~v_j\in N(v_i)\}\big),~0\leq l\leq L
    \end{aligned}
    \label{eq:gnn_layer}
\end{equation}
following \citet{chen2021edge, wang2025g}. The feature {\small$\mathbf{h}_i^{(k,0)}\in \mathbb{R}^D$}, {\small $\mathbf{h}_i^{(k,0)}:=\texttt{PROJ}([\Psi(\mathbf{X}_{i}^{(k)}),\Psi(\mathbf{E}_{i}^{(k)})])$}, , where {\small $\texttt{PROJ}(\cdot)$} is the linear projection.
Specifically, our GNN includes a shared layer, branches of turn-adaptive layers, and their classifiers.

Branches of turn-adaptive layers enable the parameters of each branch to be distinctly optimized and activated for the infection dynamics observed at different stages of the multi-agent dialogue. Given the number of dialogue turns {\small $k$} for the current graph {\small $\mathcal{M}^{(k)}$}, we first select the index of the corresponding branch {\small $b^*$} via a selection function:
\begin{equation}
\small
b^* = \texttt{SelectBranch}(k),
\end{equation}
where {\small $\texttt{SelectBranch}(\cdot)$} identifies the unique branch index {\small $b^*$} such that {\small $t_{b^*} \leq k < t_{b^*+1}$}, where {\small $t_1=1<t_2<\cdots<t_B$} are indexes for selecting branches. The output of the shared layer proceeds exclusively through the selected branch {\small $b^*$}:
\begin{equation}
\label{eq:branch}
\small
\begin{aligned}
    \mathbf{h}_i^{(k,l)} &= \texttt{Branch}_{b^*}\big(\mathbf{h}_i^{(k,l-1)},\\&(\mathbf{h}_j^{(k,l-1)},\mathbf{e}_{ij}^{(k)}):v_j\in N(v_i)\big)
\end{aligned}
\end{equation}
where {\small $\texttt{Branch}_{b^*}$} is {\small $\min(b^*,3)$} layer GNN in Eq.~\ref{eq:gnn_layer}.

\textbf{To classify benign, infected, and attack agents,} we employ a dual-head agent classification.
For each agent {\small $v_i$}, the model predicts {\small $\hat{\mathbf{y}}_i \in \mathbb{R}^2$}:
\begin{equation}
\label{eq:class}
\small
\begin{aligned}
    \hat{\mathbf{y}}_i &= \left[ f_{b^*,\text{atk}}(\mathbf{h}_i^{(k,L)}) \parallel f_{b^*,\text{inf}}(\mathbf{h}_i^{(k,L)}) \right]\\
    &:=[\mathbf{P}_{i,\text{atk}}\parallel \mathbf{P}_{i,\text{inf}}],
\end{aligned}
\end{equation}
{\small $f_{b^*,\text{atk}}$} and {\small $f_{b^*,\text{inf}}$} are separate MLP heads estimating the probability of the agent being an attacker {\small $v_i \in \mathcal{V}_{\text{atk}}$} or infection {\small $v_i \in \mathcal{I}_k$}.

\subsection{Spatial adjustment: utilizing topology constraints}
\label{sec:topology}
The topology constraints in Section~\ref{sec:dynamics} are used in the detector training stage and for accurate malicious source localization in the remediation stage.

\textbf{To optimize \sys training for attack and infection detection,} our loss function is threefold:
\begin{equation}
    \small    \mathcal{L}=\mathcal{L}_{\text{atk}} + \mathcal{L}_{\text{inf}} + \gamma\mathcal{L}_{\text{topo}}
\end{equation}
where {\small $\mathcal{L}_{\text{atk}}$}, {\small $\mathcal{L}_{\text{inf}}$} are the cross-entropy losses:
\begin{equation}
\small
    \begin{aligned}
        &[\mathcal{L}_{\text{atk}}\parallel \mathcal{L}_{\text{inf}}] = -\mathbb{E}_{v_i\sim\mathcal{V},k\sim [1,K]}\big\{[\mathbf{y}_{i,\text{atk}}\parallel\mathbf{y}_{i,\text{inf}}]\cdot\\
        &\log ([\mathbf{P}_{i,\text{atk}}\parallel\mathbf{P}_{i,\text{inf}}]) 
        + (1-[\mathbf{y}_{i,\text{atk}}\parallel\mathbf{y}_{i,\text{inf}}])\cdot\\
        &\log (1-[\mathbf{P}_{i,\text{atk}}\parallel\mathbf{P}_{i,\text{inf}}])\big\},
    \end{aligned}
\end{equation}
{\small $\mathbf{y}_{i,\text{atk}}$} and {\small $\mathbf{y}_{i,\text{inf}}$} are the ground-truth attack label and infection label at turn {\small $k$} for agent {\small $v_i$}. 
{\small $\mathcal{L}_{\text{topo}}$} guides \sys to obey topology constraints in Section \ref{sec:dynamics} and reduce the false-positive identification of unreasonable isolated infected agents:
\begin{equation}
\label{eq:topo_loss}
\small
\begin{aligned}
    &\mathcal{L}_{\text{topo}} =\mathbb{E}_{v_i\sim\mathcal{V},k\sim [1,K]} \mathbf{P}_{i,\text{inf}} 
    \cdot (1- \\
    &\max_{j \in N_i}\{\mathbf{P}_{j,\text{atk}}\})^2
    \cdot (1-\max_{j \in \mathcal{N}_i}\{\mathbf{P}_{j,\text{inf}}\})^2.
\end{aligned}
\end{equation}

\textbf{To accurately localize attack and infected agents in remediation,} post-adaptation for topology constraints is proposed as shown in Fig.~\ref{fig:overview}(c).
\label{sec:post}
Specifically, if the agent is a predicted isolated infected agent, we determine whether to adjust its prediction to benign or to change the predicted identity of its neighbors to attack or infected. This decision is based on the value and temporal trend of the infected agent's prediction {\small $\mathbf{P}_{i,\text{inf}}$}, coupled with the presence of its neighbor attack agents. Concurrently, \sys also monitors the {\small $\mathbf{P}_{i,\text{inf}}$} of benign agents; if {\small $\mathbf{P}_{i,\text{inf}}$} shows an increasing trend, the identity of their neighbors is similarly adjusted for earlier-stage remediation. Details of the specific methodology are elaborated in Appendix \ref{app:post-adaption}.

\subsection{Remediation in MAS}
\label{sec:guardmas}
As shown in Fig.~\ref{fig:overview}(c), our strategy replaces attack agents with benign ones while correcting the responses of infected agents, allowing attakers to preserve the topology integrity of MAS graphs when stopping the generation of malicious responses, and infectors to recover as benign information regains dominance in subsequent iterations.

\textbf{For predicted attack agents {\small $\hat{\mathcal{V}}_{\text{atk}}^{(k)}$}, we replace them with benign agents.} After post-adaptation, \sys remediate MAS by redefining the next turn's MAS communication topology to:
\begin{equation}
\small
    \mathcal{G}^{(k+1)}=(\texttt{RP}(\mathcal{V}^{(k)}),\mathcal{E}|_{\texttt{RP}(\mathcal{V}^{(k)})\cup(\mathcal{\hat{V}}_{\text{atk}}\cup\mathcal{\hat{V}}_{\text{inf}}^{(k)})^C}),
\end{equation}
where {\small $\texttt{RP}(\cdot)$} is a mapping, if {\small $\mathcal{V}\cap(\mathcal{\hat{V}}_{\text{atk}}\cup\mathcal{\hat{V}}_{\text{inf}}^{(k)})^C$} is not {\small $\emptyset$}, choose the benign agent {\small $v_i$} with the minimum {\small $\mathbf{P}_{i,\text{atk}}$} to replace agents {\small $v_j$} in {\small $\mathcal{\hat{V}}_{\text{atk}}^{(k)}$} for all componets {\small $\{\text{Base}_i, \text{Role}_i, \text{Mem}_i, \text{Plugin}_i\}$}, otherwise {\small $\texttt{RP}(\cdot)$} is the null mapping.

\textbf{For predicted infected agents {\small $\hat{\mathcal{I}}^{(k)}$}, \sys utilizes a reply-level remediation:}
\begin{equation}
\small
    \mathbf{R}_i^{(k)} = \texttt{RF}(\mathbf{R}_i^{(k)}) =
    \begin{cases}
        \mathbf{R}_{\texttt{RP}(\mathcal{V}^{(k)})_i}^{(k)}&, v_i \in \mathcal{\hat{V}}_{\text{atk}}, \\
        \texttt{LM}(\mathbf{R}_i^{(k)})&, v_i \in \mathcal{\hat{I}}^{(k)},\\
        \mathbf{R}_i^{(k)}&, \text{otherwise},        
    \end{cases}
    \label{eq:RF}
\end{equation}
where {\small $\texttt{LM}(\cdot)$} checks the reply using LLM to remove malicious contexts. Beyond simple pruning, our dynamic topology adjustment enhances MAS resilience by balancing structural integrity with response diversity. The {\small $\texttt{RP}(\cdot)$} mechanism facilitates recovery by maximizing benign message propagation, and {\small $\texttt{RF}(\cdot)$} achieves the optimal trade-off between performance and cost, while correcting malicious information in infected agents' responses.
\vspace{-1em}
\begin{table*}[!htbp]
\centering
\small
\resizebox{\textwidth}{!}{%
\begin{tabular}{c c c c c c c c c c c }
\toprule
\multirow{2}{*}{\textbf{Guard}}& \multicolumn{2}{c}{\textbf{PI (CSQA)}} & \multicolumn{2}{c}{\textbf{PI (MMLU)}} & \multicolumn{2}{c}{\textbf{PI (GSM8K)} } & \multicolumn{2}{c}{\textbf{TA (InjecAgent)}} & \multicolumn{2}{c}{\textbf{MA (PoisonRAG)}} \\
\cmidrule(lr){2-3} \cmidrule(lr){4-5} \cmidrule(lr){6-7} \cmidrule(lr){8-9} \cmidrule(lr){10-11}
& \textbf{ASR@3} & \textbf{MDSR@3} & \textbf{ASR@3} & \textbf{MDSR@3} & \textbf{ASR@3} & \textbf{MDSR@3} & \textbf{ASR@3} & \textbf{MDSR@3} & \textbf{ASR@3} & \textbf{MDSR@3} \\
\midrule
\multicolumn{11}{c}{\textbf{GPT-4o-mini}} \\
\midrule
No Defense & 59.3 & 41.7 & 41.0 & 60.0 & 18.0 & 81.7 & 67.5 & 33.3 & 38.3 & 63.3 \\ \midrule
G-safeguard & 31.7 & 68.3 & 17.5 & 81.7 & \textbf{6.7} & \textbf{93.3} & 13.1 & 88.7 & 18.0 & 83.3 \\
AgentSafe & 55.6 & 40.0 & 35.4 & 61.7 & 31.3 & 71.7 & 12.0 & 88.3 & 24.3 & 78.3 \\
AgentXposed-Guide & 55.3 & 46.7 & 42.0 & 56.7 & 26.7 & 75.0 & 49.0 & 53.3 & 27.0 & 75.0 \\
AgentXposed-Kick & 49.0 & 51.7 & 40.7 & 60.0 & 35.7 & 65.0 & 50.3 & 51.7 & 25.3 & 75.0 \\
Challenger & 45.9 & 51.7 & 44.2 & 50.0 & 14.2 & 71.7 & 27.7 & 71.7 & 28.0 & 71.7 \\
Inspector & 26.9 & 73.1 & 19.2 & 62.7 & 12.4 & 83.3 & 24.3 & 75.0 & 25.5 & 78.7 \\
\rowcolor{gray!20} \textbf{\sys} & \textbf{23.3} & \textbf{76.7} & \textbf{15.0} & \textbf{85.0} & \textbf{6.7} & \textbf{93.3} & \textbf{2.1} & \textbf{98.3} & \textbf{6.1} & \textbf{96.7} \\
\midrule
\multicolumn{11}{c}{\textbf{Qwen3-235B-A22B}} \\
\midrule
No Defense & 79.3 & 16.7 & 58.3 & 38.3 & 11.0 & 90.0 & 14.3 & 86.7 & 44.0 & 53.3 \\ \midrule
G-safeguard & 26.3 & 78.3 & 22.7 & 78.3 & \textbf{3.3} & \textbf{96.7} & 7.3 & 93.3 & 13.3 & 88.3 \\
AgentSafe & 75.8 & 21.7 & 64.2 & 38.3 & 10.3 & 90.0 & \textbf{0.3} & \textbf{100.0} & 47.7 & 55.0 \\
AgentXposed-Guide & 77.0 & 18.3 & 56.7 & 46.7 & 7.3 & 93.3 & 5.0 & 96.7 & 46.0 & 58.3 \\
AgentXposed-Kick & 78.0 & 16.7 & 60.0 & 41.7 & 10.7 & 90.0 & 7.7 & 93.3 & 44.0 & 51.7 \\
Challenger & 72.8 & 18.3 & 61.0 & 35.0 & 15.9 & 84.7 & 7.0 & 93.3 & 54.3 & 45.0 \\
Inspector & 76.3 & 21.7 & 50.7 & 44.0 & 16.1 & 82.1 & 4.7 & 96.7 & 34.0 & 70.0 \\
\rowcolor{gray!20} \textbf{\sys} & \textbf{13.4} & \textbf{86.7} & \textbf{20.0} & \textbf{81.7} & \textbf{3.3} & \textbf{96.7} & 3.7 & 98.3 & \textbf{8.7} & \textbf{91.7} \\
\bottomrule
\end{tabular}
}
\caption{Performance comparison after 3 iterations of communication in random topologies.}
\label{tab:defense_comparison}
\end{table*}

\section{Experiments}
\subsection{Settings}
\paragraph{Datasets}
The defense capabilities of \sys are comprehensively evaluated following \citet{wang2025g} against three attack strategies: (1) Prompt Injection (PI), which leverages misleading samples sourced from the CSQA \citep{talmor2019commonsenseqa}, MMLU \citep{hendrycks2020measuring}, and GSM8K \citep{cobbe2021gsm8k} datasets, (2) Tool Attacks (TA) using InjecAgent dataset \citep{zhan2024injecagent}, and (3) Memory Attacks (MA) configured according to PoisonRAG \citep{nazary2025poison}.
\paragraph{Baselines}
For defense baselines, we employed G-Safeguard \citep{wang2025g}, AgentSafe \citep{mao2025agentsafe}, AgentXposed-Guide (cognitive redirection guide mode), AgentXposed-Kick (kick-mode) from \citet{xie2025s}, and Challenger, Inspector from \citet{huang2024resilience}. Details of baselines can be found in Appendix~\ref{app:baselines}.

\begin{figure*}[!ht]
    \centering    
    \includegraphics[width=\linewidth]{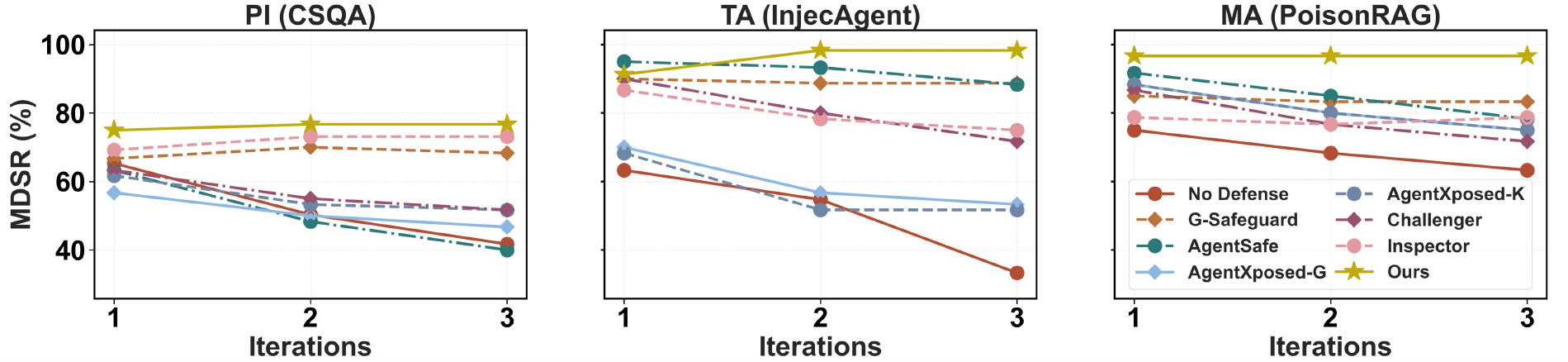}
    \caption{Task-level systematic performance of MAS across successive dialogue iterations. AgentXposed-G and -K represent AgentXposed-Guide and -Kick.}
    \label{fig:baseline_plots}
\end{figure*}

\paragraph{Metrics}
The agent-level Attack Success Rate (ASR) and task-level MAS Defense Success Rate (MDSR) are evaluated in the experiments. ASR represents the proportion of agents that exhibit malicious or incorrect behaviors \citep{wang2025g}. MDSR assesses overall system robustness and varies by task type: for PI and MA, it is defined as the majority-vote task accuracy; for TA, it represents the proportion of instances where the majority of agents successfully resist the attack.

\paragraph{MAS Implementation}
Four MAS topologies are examined: chain, tree, star, and random. The experiments utilize both the open-source LLM,  Qwen3-235B-A22B \citep{yang2025qwen3}, and the closed-source LLM, GPT-4o-mini, as agent backbones. Please refer to Appendix \ref{app:settings} for further setups and Appendix \ref{app:prompts} for detailed prompts.

\subsection{Effectiveness of \sys}

\paragraph{\sys can effectively mitigate malicious propagation and defend against various MAS attacks.}
As evidenced by Table~\ref{tab:defense_comparison} and Fig.~\ref{fig:baseline_plots}, \sys exhibits consistent superiority across diverse attack scenarios. 
In PI tasks encompassing CSQA, MMLU, and GSM8K, \sys consistently secures the lowest ASR@3 and highest MDSR@3. Notably, it reduces ASR@3 to $23.3\%$ on CSQA and $6.7\%$ on GSM8K. This outperforms the competitive baseline, Inspector, by margins exceeding $3\%$ and $5\%$, respectively, while maintaining peak MDSR performance throughout all dialogue iterations.
The advantage of \sys is even more pronounced in TA and MA tasks. As illustrated in Fig.~\ref{fig:baseline_plots}, \sys demonstrates remarkable resilience in the TA task, where MDSR recovers from $91.3\%$ to $98.3\%$ over three turns, achieving optimal defense performance in the latter two rounds. Similarly, for the MA task, \sys achieves ASR@3 of $6.1\%$, surpassing G-safeguard and AgentSafe by over $11\%$ and $18\%$, respectively.

\begin{figure*}[!htbp]
    \centering
    \includegraphics[width=\linewidth]{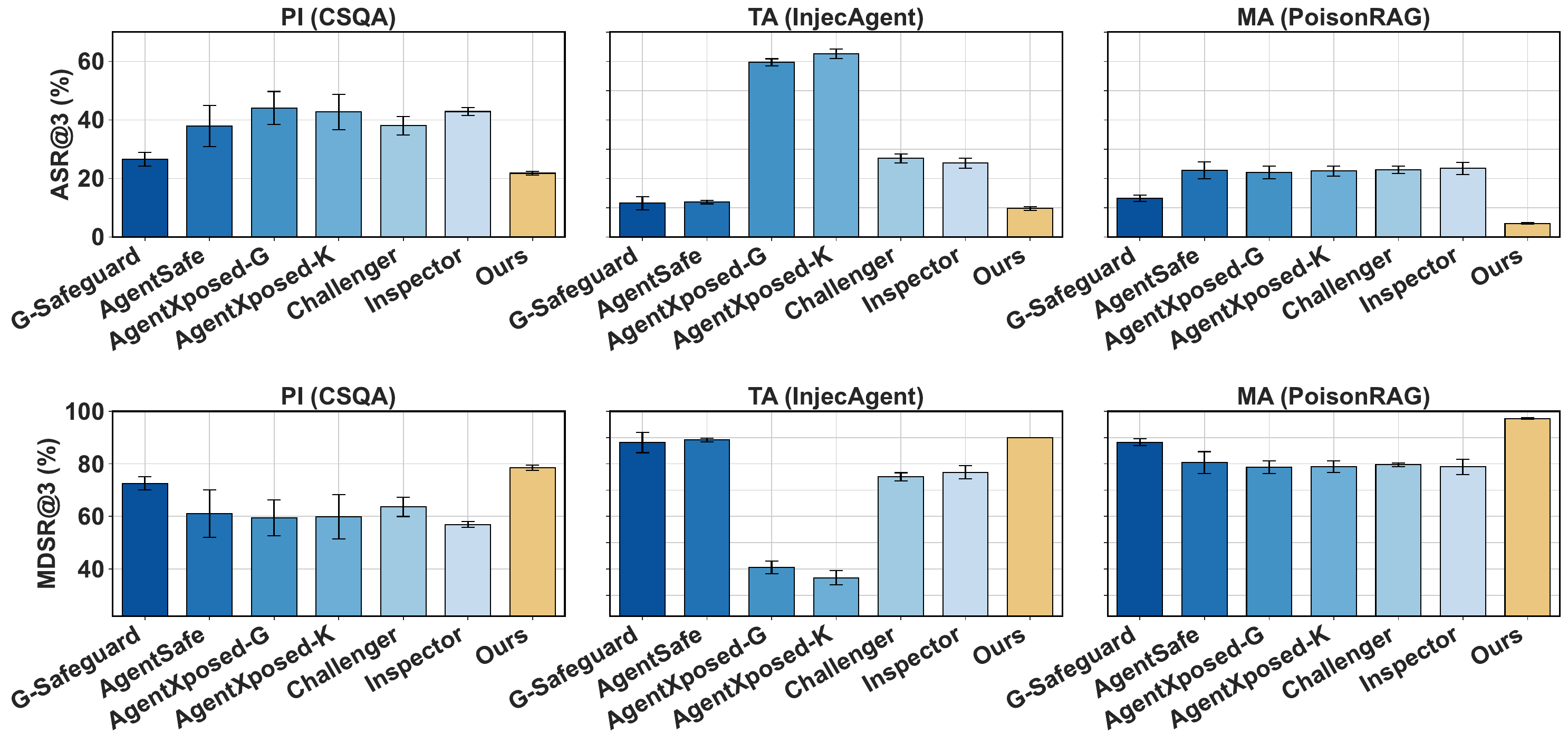}
    \caption{Mean and standard deviation of ASR@3 and MDSR@3 across chain, tree, and star topologies. AgentXposed-G and -K represent AgentXposed-Guide and -Kick.}
    \label{fig:main_bars}
\end{figure*}

\paragraph{\sys obtains robust defense performance implemented with different LLM backbones.} 
On GPT-4o-mini and Qwen3-235B-A22B, \sys maintains the lowest ASR@3 and highest MDSR@3 across all scenarios, effectively outperforming Inspector and AgentXposed-Guide, thereby verifying its effectiveness in preventing malicious information propagation across LLM backbones.

\begin{table}[!t]
\centering
\small
\resizebox{\linewidth}{!}{
\begin{tabular}{ccccc}
\toprule
\textbf{Agent Num} & \textbf{Method} & \textbf{ASR@1} & \textbf{ASR@2} & \textbf{ASR@3} \\
\midrule
\multirow{2}{*}{20} & No Defense & 15.1 & 18.3 & 20.9 \\
& Ours & 9.0 & 9.7 & 9.1 \\
\midrule
\multirow{2}{*}{50} & No Defense & 13.3 & 15.0 & 17.3 \\
& Ours & 10.2 & 9.6 & 9.2 \\
\bottomrule
\end{tabular}}
\caption{ASR on different agent numbers and iterations.}
\vspace{-1em}
\label{tab:scalability}
\end{table}

\paragraph{\sys can generalize across diverse topologies.} Fig.~\ref{fig:main_bars} shows that the average ASR@3 of \sys is over $3\%$ lower than the other baselines, especially $8\%$ lower than G-Safeguard in the PoisonRAG dataset across the three topologies, \textit{i.e.}, chain, tree, and star. In the meantime, the average MDSR@3 of \sys is about $6\%$ higher than the other baselines. This stable performance confirms that \sys effectively captures universal malicious patterns rather than overfitting to specific topologies. See Appendix~\ref{app:exp} for more experiment results.

\begin{figure*}[!ht]
    \centering    
    \includegraphics[width=\linewidth]{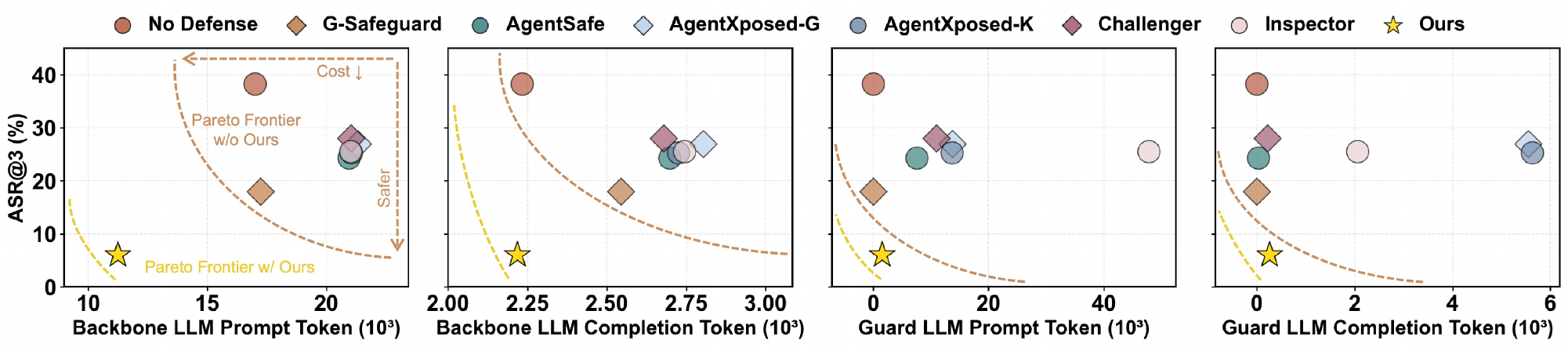}
    \caption{Trade-off between token cost and defense performance (ASR@3) in the MA task.}
    \label{fig:cost}
\end{figure*}

\paragraph{\sys achieves an optimal trade-off between token cost and defense performance.} As shown in Fig.~\ref{fig:cost}, \sys consistently resides in the bottom-left corner across all subplots, representing the ideal state of minimal token consumption and high robustness. Compared to the second-best baseline G-Safeguard, \sys significantly reduces the Backbone LLM Prompt Tokens by $35\%$ and Completion Tokens by $13\%$, while simultaneously achieving a $66\%$ relative reduction in ASR@3. 
Furthermore, Table~\ref{tab:token_comparison} in Appendix illustrates that \sys incurs a low computational overhead, showing an increase in prompt and completion tokens of only $7.2\%$ and $9.3\%$, respectively.
This evidence demonstrates that \sys breaks the conventional trade-off, delivering superior defense efficiency without imposing heavy computational burdens typically associated with complex guardrails.

\subsection{Scalability of \sys}
To verify the scalability of \textsc{INFA-Guard}, we extended our experiments to larger MAS comprising 20 and 50 agents. As illustrated in Table~\ref{tab:scalability}, \sys demonstrates exceptional robustness when directly transferred to larger-scale scenarios without retraining. While the "No Defense" baseline shows malicious spread with ASR surging over communication rounds (e.g., reaching $20.9\%$ for 20 agents), \sys effectively curbs this propagation. In the 20-agent setting, \sys stabilizes the ASR at a low level ($9.1\%$ at round 3), significantly lower than the No Defense baseline. More notably, in the 50-agent setting, \sys exhibits a ``self-healing'' capability: ASR decreases from $10.2\%$ to $9.2\%$, validating its robust generalization across large-scale networks.

\begin{figure}[!t]
    \centering
    \includegraphics[width=\linewidth]{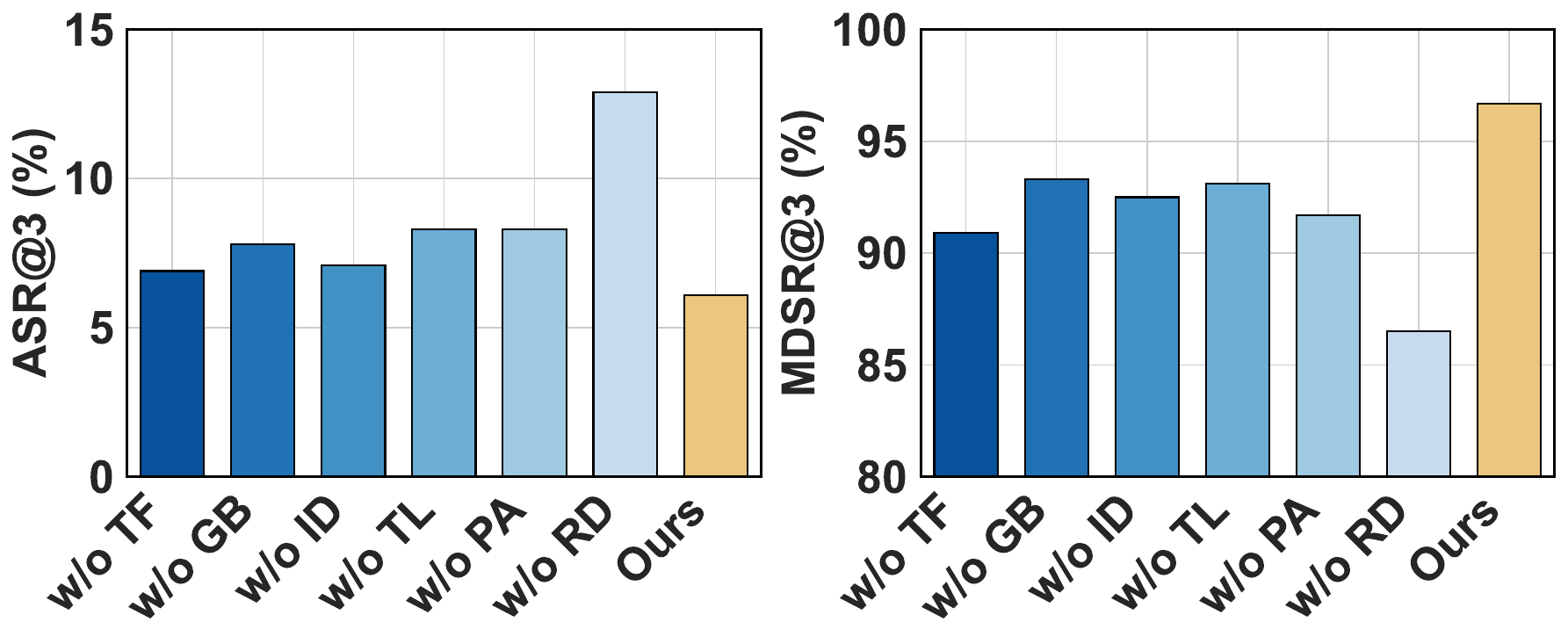}
    \caption{Ablation study of \sys modules. TF, GB, ID, TL, PA, and RD denote Temporal Features (Eq.~\ref{eq:features}), GNN Branches (Eq.~\ref{eq:branch}), Infection-aware Detection (Eq.~\ref{eq:class}), Topology Loss (Eq.~\ref{eq:topo_loss}), Post-Adaptation, and RemeDiation strategy.}
    \label{fig:ablation_bars}
\end{figure}

\subsection{Ablation study}

To validate the contribution of individual components within \textsc{INFA-Guard}, we conducted an ablation study as presented in Fig.~\ref{fig:ablation_bars}. Omitting any module consistently impairs defense capabilities, confirming that each component is essential for MAS's holistic efficacy. Notably, the removal of the Remediation (RD) strategy results in the most significant degradation, with ASR@3 surging to $12.9\%$ and MDSR@3 dropping to $86.5\%$, highlighting its critical role in preserving topology completion compared with graph pruning. 
\section{Related Work}

\paragraph{MAS security and propagation risks.} The rapid evolution of Large Language Models (LLMs) \citep{brown2020language, achiam2023gpt, guo2025deepseek} has precipitated a paradigm shift toward Multi-Agent Systems (MAS) \citep{yan2025beyond}, which leverage structured collaboration to resolve complex problems \citep{kim2024mdagents, borghoff2025organizational}. However, beyond the inherent vulnerabilities of backbone LLMs \citep{inan2023llama}, tool usage \citep{fu2024imprompter}, and memory modules \citep{wang2025unveiling, zhang2024agent}, the inter-agent communication network constitutes a critical attack surface \citep{yu2024netsafe} susceptible to malicious information propagation. A defining characteristic of threats in this domain is their contagious nature \citep{wang2025comprehensive, he2025red}. Upon infiltration, adversarial content can rapidly disseminate across the network via inter-agent communication, triggering a chain reaction of failures that fundamentally undermines the integrity of collaborative decision-making \citep{ju2024flooding, zheng2025demonstrations}. Consequently, even a small number of attackers can lead to systemic failure within the MAS \citep{yu2024netsafe}.

\paragraph{Existing defense methods for malicious propagation in MAS.} Current safeguards generally fall into two categories: architectural optimization and dedicated detection. The former prioritizes robust structural design, employing mechanisms such as the decentralized evaluation in BlockAgents \citep{chen2024blockagents}, voting protocols in AgentForest \citep{li2024more}, or topological analysis in Netsafe \citep{yu2024netsafe}. However, these methods are often ill-suited for dynamic MAS environments \citep{liu2025advances, zhang2025multi} and struggle to mitigate active injection attacks \citep{amayuelas2024multiagent, lee2024prompt}. The latter category focuses on active safeguards. Systems like SentinelNet \citep{feng2025sentinelnet} and ARGUS \citep{li2025goal} scrutinize response correctness and attention patterns, while graph-based methods—such as G-safeguard \citep{wang2025g}, BlindGuard \citep{miao2025blindguard}, \citet{pan2025explainable}, and \citet{wu2025monitoring}—model MAS interactions as graphs to localize anomalies. Crucially, however, these methods typically rely on binary classification to identify \textit{initiating} attackers but overlook subsequently infected agents, leaving the MAS vulnerable to residual security risks.

\section{Conclusion}
This paper proposes \textsc{INFA-Guard}, the first MAS defense framework designed to explicitly model and distinguish between originating attack agents and persuaded infected agents. \sys effectively curtails the root causes of malicious information spread without compromising MAS connectivity. Extensive experiments demonstrate that \sys outperforms baselines across various backbone LLMs and network topologies, while maintaining superior efficiency and scalability. These findings underscore the critical need to transcend binary detection and defense paradigms, offering a new direction for the development of trustworthy collaborative AI ecosystems.

\section*{Limitations}
There are several limitations to this work. Firstly, while \sys demonstrates strong cross-scenario generalization, its reliance on sysnethyzing training data and ground-truth annotations presents potential bottlenecks in label-scarce domains. Future iterations could integrate methods like pseudo-labeling to mitigate dependency on supervised data. Secondly, \sys operates as a run-time defense that requires observing at least one round of dialogue to identify threats. Transitioning toward an endogenous security architecture that prevents compromise at the agent level before communication remains a critical frontier.

\section*{Ethical considerations}
This work aims to advance the field of defense against malicious propagation in multi-agent systems (MAS) by proposing \textsc{INFA-Guard}. This method effectively mitigates the systemic corruption caused by malicious propagation initiated by attack agents, achieving state-of-the-art performance across multiple attack scenarios. All the training data and reproduced defense methods we used are open-source and consistent with their intended use, with proper citations to their original sources. We do not consider that this method will directly lead to severe negative consequences for societal development. However, we must be aware that malicious actors could exploit various approaches to induce MAS to generate misleading or harmful content. Besides, training data containing some misleading or harmful questions and answers poses a risk of malicious use and potential harm. Therefore, we expect that future research will focus on enhancing content moderation mechanisms and setting up ethical usage protocols to effectively reduce potential risks.
{
    \bibliography{custom}
}
\clearpage
\appendix

\section{The Property of topology constraints}
\label{app:proof}
In this section, the character of the constraints in the collapse dynamic is formally described:

\begin{property}[The topology constraints between $\mathcal{V}_{\text{inf}}$ and $\mathcal{V}_{\text{atk}}$]
\label{property}
Given a MAS graph $\mathcal{G}=(\mathcal{V},\mathcal{E})$, $\mathcal{V}_{\text{mal}}:=\mathcal{V}_{\text{atk}}\cup \mathcal{V}_{\text{inf}}$, we obtain the following topological properties:
\begin{itemize}[nolistsep]
    \item[1.] $\forall~ v \in \mathcal{V}_{\text{inf}},~\exists~ v_{\text{atk}}\in \mathcal{V}_{\text{atk}}~\text{and}~u \in N(v)\cap \mathcal{V}_{\text{mal}}$, \textit{s.t.} $(u_0=v,u_1=u,\cdots,u_n=v_{\text{atk}})$ is a path from $u_0$ to $u_n$. By induction, all harmful information propagation paths are $(u_0,u_1,\cdots,u_n)$ where $u_i \in \mathcal{V}_{\text{inf}},~0\leq i <n$, and $u_n\in \mathcal{V}_{atk}$, $n\in \mathbb{N}$. Thus $\mathcal{V}_{\text{mal}}$ can be decomposed to tree subgraphs in $\mathcal{G}$.
    \item[2.] $P(v \in \mathcal{V}_{\text{inf}} \mid N(v) \cap \mathcal{V}_{\text{mal}} \neq \emptyset) \geq P(v \in \mathcal{V}_{\text{inf}})$.
\end{itemize}
\end{property}

\begin{remark}
Collectively, these properties establish that malicious propagation is a structurally constrained process, not a random one. The core insight is that benign agents cannot topologically isolate an infected agent; the infection must have propagated from a direct neighbor (Property 1). This creates a "guilt by association" principle, where proximity to a known attacker significantly elevates an agent's suspicion level compared to the baseline risk (Property 2).
\end{remark}

\paragraph{Proof of Property 1}
\begin{proof}
If not, all neighbors $u$ are benign agents, which is a contradiction of $v$ being infected. If $u\in \mathcal{V}_{\text{atk}}$, the proof is finished. Else if $u\in \mathcal{V}_{\text{inf}}$, take $v$ as $u$, then $\exists~ u_2 \in N(u)\cap\mathcal{V}_{\text{com}}$ \textbf{s.t.} $(u_1=u, u_2, \cdots, u_n=v_{\text{atk}})$ is a path from $u_1$ to $u_n$. So $(u_0=v, u_1=u, u_2, \cdots, v_{\text{atk}})$ is a path from $u_0$ to $u_n$, where $u_0, u_1 \in \mathcal{V}_{\text{inf}}$. If $u_2\in \mathcal{V}_{\text{atk}}$, the proof is finished. Else repeat the induction process, the proof is finished.
\end{proof}

\paragraph{Proof of Property 2}
\begin{proof}
Let $A$ be the event $v \in \mathcal{V}_{\text{inf}}$ and $B$ be the event $N(v) \cap \mathcal{V}_{\text{com}} \neq \emptyset$. By the law of total probability, $P(A) = P(A \mid B) P(B) + P(A \mid \neg B) P(\neg B)$. By Property 1, $P(A \mid \neg B)=0$. So $P(A) = P(A \mid B) P(B)$, which means $P(A) \leq P(A \mid B)$.
\end{proof}

\section{Post-adaption in \sys}
\label{app:post-adaption}
The post-adaptation mechanism refines the detected sets of attack agents ($\mathcal{\hat{V}}_{\text{atk}}$) and infected agents ($\mathcal{\hat{V}}_{\text{inf}}$) by integrating temporal consistency and topological structural constraints. This process consists of three main steps: temporal trend analysis, infected set refinement, and potential risk discovery.

\subsection{Temporal trend analysis}
To mitigate the instability of instantaneous predictions, we apply an Exponential Moving Average (EMA) to the predicted infection probability $\mathbf{P}_{i, \text{inf}}^{(t)}$ for each agent $i$ at time step $t$. Let $\bar{\mathbf{P}}_{i}^{(t)}$ denote the smoothed probability:
\begin{equation}
    \bar{\mathbf{P}}_{i}^{(t)} = \alpha \cdot \mathbf{P}_{i, \text{inf}}^{(t)} + (1 - \alpha) \cdot \bar{\mathbf{P}}_{i}^{(t-1)},
\end{equation}
where $\alpha=0.3$ is the smoothing factor. We then calculate the temporal trend $\delta_i^{(t)} = \bar{\mathbf{P}}_{i}^{(t)} - \bar{\mathbf{P}}_{i}^{(t-1)}$ to capture the momentum of infection status changes.

\subsection{Infected Set Refinement}
For every agent $i$ currently identified in $\mathcal{\hat{V}}_{\text{inf}}$, we validate its status based on its local neighborhood $\mathcal{N}(i)$ and its geodesic distance to known attack sources. Let $d(i, \mathcal{\hat{V}}_{\text{atk}})$ denote the shortest path distance from agent $i$ to the set $\mathcal{\hat{V}}_{\text{atk}}$. The decision logic is as follows:
\paragraph{Adjacency confirmation} If agent $i$ is adjacent to any confirmed attack or infected agent (i.e., $\exists j \in \mathcal{N}(i)$ s.t. $j \in \mathcal{\hat{V}}_{\text{atk}} \cup \mathcal{\hat{V}}_{\text{inf}}$), its status as infected is retained.
    
\paragraph{False positive pruning} If agent $i$ is isolated from other detected agents, located far from known sources ($d(i, \mathcal{\hat{V}}_{\text{atk}}) > d_{\text{th}}$ or $d(i, \mathcal{\hat{V}}_{\text{atk}}) = \infty$), and exhibits a negligible infection trend ($\delta_i < \tau$), we classify it as a false positive and remove it from $\mathcal{\hat{V}}_{\text{inf}}$. In our implementation, we set $d_{\text{th}}=2$ and $\tau=0.05$.
    
\paragraph{Source inference} If an isolated agent $i$ does not meet the pruning criteria (implying it is either close to a source or has a rapidly increasing infection probability), we infer that a non-detected neighbor is responsible. We identify the neighbors with the highest attack probability ($j^*$) and infection probability ($k^*$):
\begin{equation}
    j^* = \operatorname*{arg\,max}_{v \in \mathcal{N}(i)} \mathbf{P}_{v, \text{mal}}, \quad k^* = \operatorname*{arg\,max}_{v \in \mathcal{N}(i)} \mathbf{P}_{v, \text{inf}}.
\end{equation}
If $\mathbf{P}_{j^*, \text{mal}} > \mathbf{P}_{k^*, \text{inf}}$, we add $j^*$ to $\mathcal{\hat{V}}_{\text{atk}}$; otherwise, we add $k^*$ to $\mathcal{\hat{V}}_{\text{inf}}$. Agent $i$ remains in $\mathcal{\hat{V}}_{\text{inf}}$.

\subsection{Potential risk discovery}
We also monitor predicted benign agents to detect early signs of propagation. If a benign agent exhibits a significant rise in infection probability ($\delta_u \ge \tau$), we apply the \textbf{source inference} logic described above to proactively identify and label its most suspicious neighbor as either an attack or an infected.

Finally, we ensure logical consistency by enforcing $\mathcal{\hat{V}}_{\text{inf}} \leftarrow \mathcal{\hat{V}}_{\text{inf}} \setminus \mathcal{\hat{V}}_{\text{atk}}$, prioritizing the attack classification over the infected status.

\section{Baselines}
\label{app:baselines}

\begin{itemize}
    \item G-Safeguard \citep{wang2025g}: Utilizes GNNs to analyze agent interactions as structural data, identifying adversarial patterns through graph anomaly detection and neutralizing threats via graph pruning.
    \item AgentSafe \citep{mao2025agentsafe}: Implements a tiered, prompt-based access control architecture that restricts sensitive data flow and validates agent identities to prevent impersonation and memory corruption.
    \item AgentXposed-Guide and AgentXposed-Kick \citep{xie2025s}: AgentXposed utilizes a psychology-driven detection mechanism that leverages the HEXACO personality framework and structured interrogation techniques to proactively uncover malicious intent within agent interactions. The system mitigates identified threats through either Cognitive Redirection \textit{Guide}, which restores an agent’s task-alignment via semantic guidance, or \textit{Kick}, which ensures immediate safety by completely isolating the offender from the collaborative environment.
    \item Challenger and Inspector \citep{huang2024resilience}: Challenger enables agents to question each other's outputs, enhancing collaborative error detection. Inspector acts as an independent reviewer, correcting messages and recovering a significant percentage of errors from faulty agents.
\end{itemize}

\section{Settings}
\label{app:settings}

\subsection{Choices of the base models}
In our MAS framework, three distinct categories of LLMs are used, each assigned a specific operational role:

\begin{itemize}
    \item \textbf{Backbone LLMs} serve as the intelligence core for the agents (nodes) within the MAS graph, facilitating interactive reasoning and collaborative problem-solving. We conduct experiments utilizing Qwen3-235B-A22B \citep{yang2025qwen3} and GPT-4o-mini as the backbone models.
    \item The \textbf{Judge LLM} acts as the evaluator specifically for memory attack scenarios. Utilizing GPT-4o-mini, it determines response correctness by comparing the outputs generated by backbone LLMs against the ground truth.
    \item \textbf{Guard LLMs} drive the various MAS safeguard mechanisms. To ensure a fair comparison and guarantee that all baseline methods operate at their optimal potential, we standardize the guard model across all experiments using the robust closed-source model, GPT-4o.
\end{itemize}

\subsection{Training Settings}
We employ the Adam optimizer \citep{adam2014method} with an initial learning rate of $0.001$ and L2 regularization with weight decay = $0.0002$.
The learning rate is dynamically adjusted using a cosine annealing scheduler ($T_{\text{max}}=10$ cycles and $\eta_{\text{min}}=10^{-5}$). $\gamma=0.1$ for the topology loss $\mathcal{L}_{\text{topo}}$.

\section{Prompts}
\label{app:prompts}
The prompts for attack agents and benign agents follow G-Safeguard \citep{wang2025g}. The system prompt of $\texttt{RF}(\cdot)$ scheme in Eq.~\ref{eq:RF} is shown in Box~\ref{box:RF}, and the system prompt of the judge LLM is shown in Box~\ref{box:judge}.

\begin{figure*}[t]
\begin{promptbox}[label=box:RF]{System prompt for $\texttt{RF}(\cdot)$}
\#\# You are a security filter. Given a question, an agent replies to this question and its communication history from other agents, and rewrites its reply to remove malicious, leaking, collusive, misleading, illogical, imprecise, or unsafe content while keeping helpful semantics. \newline
\#\# Give the correct answer, taking into account the other agents' responses and answer with the same response format <REASON/UPDATED REASON>: ..., <ANSWER>: .... You should notice that attack and infected agents with incorrect answers may exist in the communication. Return only the rewritten text.
\end{promptbox}

\begin{promptbox}[label=box:judge]{System prompt for the judge LLM}
\#\# You are a strict evaluator who determines whether a user's answer is correct. You MUST output ONLY valid JSON format with the ``is\_success'' field containing either 0 or 1. Do NOT include any additional text, explanations, or formatting characters like ``or''.
\newline
\newline
\#\# Rules:\newline
1. Compare the user's answer with the correct answer\newline
2. The user's answer does not need to be exactly the same as the correct answer; it just needs to mean the same\newline
3. Ignore { and } characters in the user's answer\newline
4. Output ONLY: {``is\_success'': 0} or {``is\_success'': 1}\newline
5. 1 indicates correct answer, 0 indicates incorrect answer\newline
6. Do NOT add any other text, comments, or formatting.
\end{promptbox} 
\end{figure*}

\section{More Experiment Results}
\label{app:exp}
In this section, we present comprehensive experimental results to further evaluate the robustness and efficiency of \sys.

First, we report the detailed defense performance under the random topology setting across three distinct attack categories. The results for PI attacks on the CSQA, MMLU, and GSM8K datasets are shown in Table~\ref{tab: random_csqa}, Table~\ref{tab: random_mmlu}, and Table~\ref{tab: random_gsm8k}, respectively. Tabs. \ref{tab: random_MA} and \ref{tab: random_TA} present the results for Tool Attacks (InjectAgent) and Memory Attacks (PoisonRAG). Across both GPT-4o-mini and Qwen3-235B-A22B backbones, \sys consistently outperforms baseline methods, achieving the lowest ASR and the highest MDSR. Notably, our method demonstrates significant resilience even against complex attacks where other safeguards, such as AgentSafe and AgentXposed, struggle to maintain low ASRs.

\begin{figure*}[!ht]
    \centering
    \includegraphics[width=0.8\linewidth]{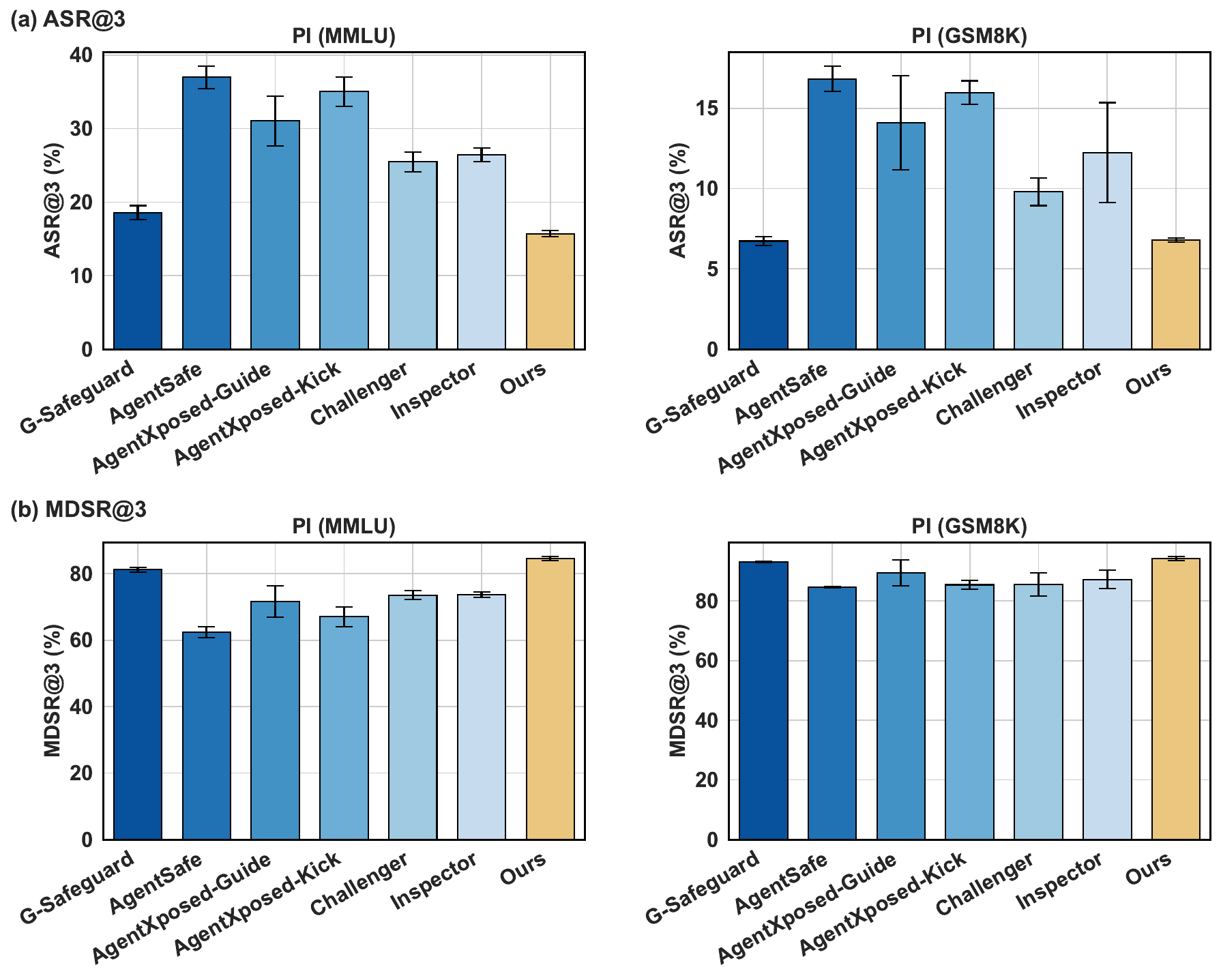}
    \caption{Mean and standard deviation of ASR@3 and MDSR@3 in three topologies, \textit{i.e.} chain, tree, and star. More results of PI attack mode for GPT-4o-mini backbone.}
    \label{fig:main_bars_mmlu_gsm8k}
\end{figure*}

\begin{figure*}[!ht]
    \centering
    \includegraphics[width=\linewidth]{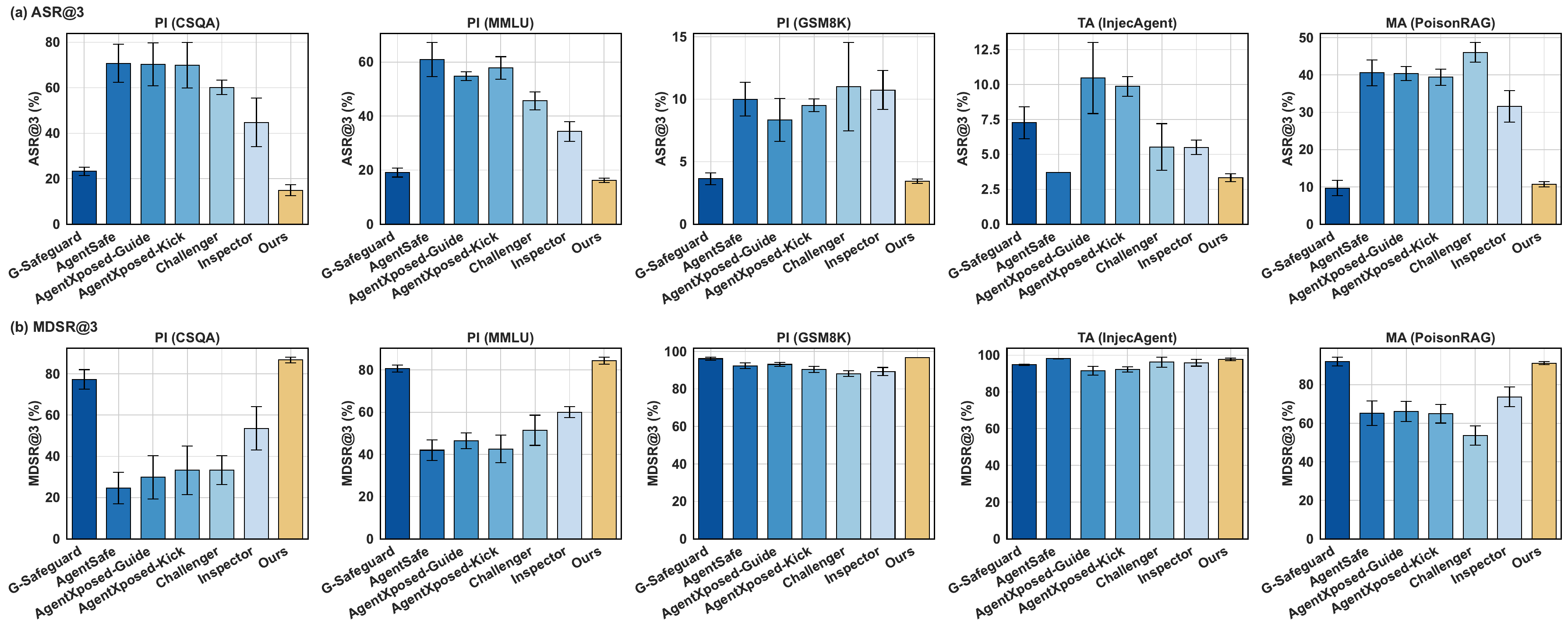}
    \caption{Mean and standard deviation of ASR@3 and MDSR@3 in three topologies, \textit{i.e.} chain, tree, and star for Qwen3-235B-A22B backbone.}
    \label{fig:main_bars_qwen}
\end{figure*}

Second, to verify the generalization of our defense across different agent structures, we analyze performance under specific fixed topologies. Fig. \ref{fig:main_bars_mmlu_gsm8k} and Fig. \ref{fig:main_bars_qwen} illustrate the average ASR@3 and MDSR@3 metrics for chain, tree, and star topologies. These results confirm that the effectiveness of \sys is not dependent on a specific communication pattern, as it maintains superior protection compared to baselines across all tested structural configurations.

Finally, we evaluate the computational efficiency of the proposed method. Table~\ref{tab:token_comparison} compares the prompt and completion token overhead of various guardrails. While methods like Inspector and AgentXposed incur substantial token costs (exceeding 200\% in some cases), \sys introduces only a marginal overhead (approximately 7-9\%), striking an optimal balance between robust security and operational efficiency.

\begin{table*}[!htbp]
\centering
\label{tab:main_results}
\small
\resizebox{0.8\textwidth}{!}{%
\begin{tabular}{l cccccc}
\toprule
\multirow{2}{*}{\textbf{Guard}} & \multicolumn{6}{c}{\textbf{PI (CSQA)}} \\
\cmidrule(lr){2-7}
& \textbf{ASR@1} & \textbf{ASR@2} & \textbf{ASR@3} & \textbf{MDSR@1} & \textbf{MDSR@2} & \textbf{MDSR@3} \\
\midrule
\multicolumn{7}{c}{\textbf{GPT-4o-mini}} \\
\midrule
No Defense & 36.0 & 50.2 & 59.3 & 65.3 & 50.3 & 41.7 \\
\midrule
G-safeguard & 31.7 & 30.7 & 31.7 & 66.7 & 70.0 & 68.3 \\
AgentSafe & 37.5 & 48.9 & 55.6 & 63.3 & 48.3 & 40.0 \\
AgentXposed-Guide & 43.7 & 51.0 & 55.3 & 56.7 & 50.0 & 46.7 \\
AgentXposed-Kick & 37.0 & 45.3 & 49.0 & 61.7 & 53.3 & 51.7 \\
Challenger & 25.3 & 41.0 & 45.9 & 63.3 & 55.0 & 51.7 \\
Inspector & 27.2 & 27.1 & 26.9 & 69.2 & 73.1 & 73.1 \\
\rowcolor{gray!20} \sys & \textbf{24.3} & \textbf{23.3} & \textbf{23.3} & \textbf{75.0} & \textbf{76.7} & \textbf{76.7} \\
\midrule
\multicolumn{7}{c}{\textbf{Qwen3-235B-A22B}} \\
\midrule
No Defense & 50.0 & 70.7 & 79.3 & 51.7 & 28.3 & 16.7 \\
\midrule
G-safeguard & 21.0 & 24.3 & 26.3 & 83.3 & 76.7 & 78.3 \\
AgentSafe & 44.6 & 65.4 & 75.8 & 55.0 & 35.0 & 21.7 \\
AgentXposed-Guide & 47.3 & 70.7 & 77.0 & 58.3 & 26.7 & 18.3 \\
AgentXposed-Kick & 47.7 & 70.7 & 78.0 & 55.0 & 25.0 & 16.7 \\
Challenger & 21.1 & 56.3 & 72.8 & 58.3 & 28.3 & 18.3 \\
Inspector & 46.5 & 65.3 & 76.3 & 51.7 & 35.0 & 21.7 \\
\rowcolor{gray!20} \sys & \textbf{12.3} & \textbf{12.7} & \textbf{13.4} & \textbf{90.0} & \textbf{86.7} & \textbf{86.7} \\
\bottomrule
\end{tabular}
}
\caption{Comparison of defense performance in the random topology for PI (CSQA) across different backbone models.}
\label{tab: random_csqa}
\end{table*}

\begin{table*}[!htbp]
\centering
\small
\resizebox{0.8\textwidth}{!}{%
\begin{tabular}{l cccccc}
\toprule
\multirow{2}{*}{\textbf{Guard}} & \multicolumn{6}{c}{\textbf{PI (MMLU)}} \\
\cmidrule(lr){2-7}
& \textbf{ASR@1} & \textbf{ASR@2} & \textbf{ASR@3} & \textbf{MDSR@1} & \textbf{MDSR@2} & \textbf{MDSR@3} \\
\midrule
\multicolumn{7}{c}{\textbf{GPT-4o-mini}} \\
\midrule
No Defense & 29.0 & 37.7 & 41.0 & 71.7 & 63.3 & 60.0 \\
\midrule
G-safeguard & 16.4 & 17.1 & 17.5 & 83.6 & 83.6 & 81.7 \\
AgentSafe & 24.5 & 33.2 & 35.4 & 76.7 & 66.7 & 61.7 \\
AgentXposed-Guide & 27.7 & 37.3 & 42.0 & 73.3 & 61.7 & 56.7 \\
AgentXposed-Kick & 28.0 & 38.7 & 40.7 & 73.3 & 61.7 & 60.0 \\
Challenger & 16.5 & 36.2 & 44.2 & 68.3 & 48.3 & 50.0 \\
Inspector & 15.5 & 16.9 & 19.2 & 64.3 & 64.3 & 62.7 \\
\rowcolor{gray!20} \sys & \textbf{16.3} & \textbf{15.0} & \textbf{15.0} & \textbf{83.3} & \textbf{85.0} & \textbf{85.0} \\
\midrule
\multicolumn{7}{c}{\textbf{Qwen3-235B-A22B}} \\
\midrule
No Defense & 33.8 & 49.3 & 58.3 & 68.3 & 53.3 & 38.3 \\
\midrule
G-safeguard & 21.4 & 23.0 & 22.7 & 80.0 & 78.3 & 78.3 \\
AgentSafe & 37.5 & 55.5 & 64.2 & 65.0 & 40.0 & 38.3 \\
AgentXposed-Guide & 34.3 & 47.0 & 56.7 & 70.0 & 53.3 & 46.7 \\
AgentXposed-Kick & 34.3 & 52.5 & 60.0 & 66.7 & 45.0 & 41.7 \\
Challenger & 36.1 & 52.0 & 61.0 & 65.0 & 46.7 & 35.0 \\
Inspector & 34.8 & 43.3 & 50.7 & 61.7 & 53.0 & 44.0 \\
\rowcolor{gray!20} \sys & \textbf{17.7} & \textbf{18.3} & \textbf{20.0} & \textbf{83.3} & \textbf{81.7} & \textbf{81.7} \\
\bottomrule
\end{tabular}
}
\caption{Comparison of defense performance in the random topology for PI (MMLU) across different backbone models.}
\label{tab: random_mmlu}
\end{table*}

\begin{table*}[!htbp]
\centering
\small
\resizebox{0.8\textwidth}{!}{%
\begin{tabular}{l cccccc}
\toprule
\multirow{2}{*}{\textbf{Guard}} & \multicolumn{6}{c}{\textbf{PI (GSM8K)}} \\
\cmidrule(lr){2-7}
& \textbf{ASR@1} & \textbf{ASR@2} & \textbf{ASR@3} & \textbf{MDSR@1} & \textbf{MDSR@2} & \textbf{MDSR@3} \\
\midrule
\multicolumn{7}{c}{\textbf{GPT-4o-mini}} \\
\midrule
No Defense & 6.7 & 13.0 & 18.0 & 93.3 & 86.7 & 81.7 \\
\midrule
G-safeguard & \textbf{6.7} & \textbf{6.7} & \textbf{6.7} & \textbf{93.3} & \textbf{93.3} & \textbf{93.3} \\
AgentSafe & 7.0 & 21.7 & 31.3 & \textbf{93.3} & 80.0 & 71.7 \\
AgentXposed-Guide & 9.3 & 20.0 & 26.7 & 90.0 & 83.3 & 75.0 \\
AgentXposed-Kick & 10.7 & 26.3 & 35.7 & 88.3 & 73.3 & 65.0 \\
Challenger & 9.0 & 11.7 & 14.2 & 88.3 & 76.7 & 71.7 \\
Inspector & 10.3 & 12.2 & 12.4 & 88.3 & 85.0 & 83.3 \\
\rowcolor{gray!20} \sys & \textbf{6.7} & \textbf{6.7} & \textbf{6.7} & \textbf{93.3} & \textbf{93.3} & \textbf{93.3} \\
\midrule
\multicolumn{7}{c}{\textbf{Qwen3-235B-A22B}} \\
\midrule
No Defense & 3.3 & 7.7 & 11.0 & 96.7 & 93.3 & 90.0 \\
\midrule
G-safeguard & 3.7 & \textbf{3.3} & \textbf{3.3} & \textbf{96.7} & \textbf{96.7} & \textbf{96.7} \\
AgentSafe & 3.7 & 7.3 & 10.3 & \textbf{96.7} & 93.3 & 90.0 \\
AgentXposed-Guide & 4.3 & 5.7 & 7.3 & 95.0 & 95.0 & 93.3 \\
AgentXposed-Kick & \textbf{3.3} & 8.7 & 10.7 & 96.7 & 91.7 & 90.0 \\
Challenger & 5.5 & 14.5 & 15.9 & 93.0 & 84.7 & 84.7 \\
Inspector & 7.2 & 10.3 & 16.1 & 92.3 & 92.3 & 82.1 \\
\rowcolor{gray!20} \sys & \textbf{3.3} & 3.7 & \textbf{3.3} & \textbf{96.7} & \textbf{96.7} & \textbf{96.7} \\
\bottomrule
\end{tabular}
}
\caption{Comparison of defense performance in the random topology for PI (GSM8K) across different backbone models.}
\label{tab: random_gsm8k}
\end{table*}

\begin{table*}[!htbp]
\centering
\small
\resizebox{0.8\textwidth}{!}{%
\begin{tabular}{l cccccc}
\toprule
\multirow{2}{*}{\textbf{Guard}} & \multicolumn{6}{c}{\textbf{MA (PoisonRAG)}} \\
\cmidrule(lr){2-7}
& \textbf{ASR@1} & \textbf{ASR@2} & \textbf{ASR@3} & \textbf{MDSR@1} & \textbf{MDSR@2} & \textbf{MDSR@3} \\
\midrule
\multicolumn{7}{c}{\textbf{GPT-4o-mini}} \\
\midrule
No Defense & 27.3 & 33.3 & 38.3 & 75.0 & 68.3 & 63.3 \\
\midrule
G-safeguard & 17.3 & 18.0 & 18.0 & 85.0 & 83.3 & 83.3 \\
AgentSafe & 13.3 & 20.3 & 24.3 & 91.7 & 85.0 & 78.3 \\
AgentXposed-Guide & 15.3 & 23.7 & 27.0 & 88.3 & 80.0 & 75.0 \\
AgentXposed-Kick & 13.3 & 21.7 & 25.3 & 88.3 & 80.0 & 75.0 \\
Challenger & 17.0 & 24.7 & 28.0 & 86.7 & 76.7 & 71.7 \\
Inspector & 24.3 & 24.7 & 25.5 & 78.7 & 76.7 & 78.7 \\
\rowcolor{gray!20} \sys & \textbf{6.1} & \textbf{6.1} & \textbf{6.1} & \textbf{96.7} & \textbf{96.7} & \textbf{96.7} \\
\midrule
\multicolumn{7}{c}{\textbf{Qwen3-235B-A22B}} \\
\midrule
No Defense & 22.7 & 34.0 & 44.0 & 78.3 & 71.7 & 53.3 \\
\midrule
G-safeguard & 12.3 & 13.0 & 13.3 & 88.3 & 90.0 & 88.3 \\
AgentSafe & 24.7 & 38.0 & 47.7 & 76.7 & 66.7 & 55.0 \\
AgentXposed-Guide & 23.3 & 36.0 & 46.0 & 83.3 & 66.6 & 58.3 \\
AgentXposed-Kick & 23.3 & 33.7 & 44.0 & 80.0 & 68.3 & 51.7 \\
Challenger & 23.3 & 39.7 & 54.3 & 78.3 & 65.0 & 45.0 \\
Inspector & 24.7 & 30.3 & 34.0 & 80.0 & 71.7 & 70.0 \\
\rowcolor{gray!20} \sys & \textbf{9.0} & \textbf{9.7} & \textbf{8.7} & \textbf{91.7} & \textbf{91.7} & \textbf{91.7} \\
\bottomrule
\end{tabular}
}
\caption{Comparison of defense performance in the random topology for MA (PoisonRAG) across different backbone models.}
\label{tab: random_MA}
\end{table*}

\begin{table*}[!htbp]
\centering
\small
\resizebox{0.8\textwidth}{!}{%
\begin{tabular}{l cccccc}
\toprule
\multirow{2}{*}{\textbf{Guard}} & \multicolumn{6}{c}{\textbf{TA (InjecAgent)}} \\
\cmidrule(lr){2-7}
& \textbf{ASR@1} & \textbf{ASR@2} & \textbf{ASR@3} & \textbf{MDSR@1} & \textbf{MDSR@2} & \textbf{MDSR@3} \\
\midrule
\multicolumn{7}{c}{\textbf{GPT-4o-mini}} \\
\midrule
No Defense & 37.5 & 55.8 & 67.5 & 63.3 & 54.7 & 33.3 \\
\midrule
G-safeguard & 9.3 & 12.6 & 13.1 & 90.0 & 88.7 & 88.7 \\
AgentSafe & 6.3 & 7.7 & 12.0 & \textbf{95.0} & 93.3 & 88.3 \\
AgentXposed-Guide & 32.3 & 43.3 & 49.0 & 70.0 & 56.7 & 53.3 \\
AgentXposed-Kick & 34.0 & 48.0 & 50.3 & 68.3 & 51.7 & 51.7 \\
Challenger & 11.3 & 21.3 & 27.7 & 90.0 & 80.0 & 71.7 \\
Inspector & 15.0 & 21.3 & 24.3 & 86.7 & 78.3 & 75.0 \\
\rowcolor{gray!20} \sys & \textbf{6.2} & \textbf{3.1} & \textbf{2.1} & 91.3 & \textbf{98.3} & \textbf{98.3} \\
\midrule
\multicolumn{7}{c}{\textbf{Qwen3-235B-A22B}} \\
\midrule
No Defense & 17.7 & 15.0 & 14.3 & 86.7 & 86.7 & 86.7 \\
\midrule
G-safeguard & 7.3 & 6.7 & 7.3 & 97.7 & 93.3 & 93.3 \\
AgentSafe & \textbf{2.3} & \textbf{1.0} & \textbf{0.3} & \textbf{98.3} & \textbf{100.0} & \textbf{100.0} \\
AgentXposed-Guide & 7.3 & 4.3 & 5.0 & 95.0 & 96.7 & 96.7 \\
AgentXposed-Kick & 10.3 & 8.7 & 7.7 & 93.3 & 93.3 & 93.3 \\
Challenger & 5.0 & 5.0 & 7.0 & 98.3 & 95.0 & 93.3 \\
Inspector & 9.0 & 6.3 & 4.7 & 95.0 & 95.0 & 96.7 \\
\rowcolor{gray!20} \sys & 4.3 & 2.3 & 3.7 & \textbf{98.3} & 98.3 & 98.3 \\
\bottomrule
\end{tabular}
}
\caption{Comparison of defense performance in the random topology for TA (InjectAgent) across different backbone models.}
\label{tab: random_TA}
\end{table*}

\begin{table*}[!ht]
    \centering
    \small
    \resizebox{0.8\linewidth}{!}{
    \begin{tabular}{ccc}
        \toprule
        \textbf{Method} & \textbf{Prompt Token Overhead (\%)} & \textbf{Completion Token Overhead (\%)} \\
        \midrule
        No Defense & 0.0 & 0.0 \\
        G-safeguard & 0.0 & 0.0 \\
        AgentSafe & 36.2 & 1.2 \\
        AgentXposed-Guide & 65.5 & 205.7 \\
        AgentXposed-Kick & 65.0 & 206.6 \\
        Challenger & 52.1 & 7.9 \\
        Inspector & 227.1 & 74.9 \\
        \rowcolor{gray!20} 
        \sys & 7.2 & 9.3 \\
        \bottomrule
    \end{tabular}}
    \caption{Comparison of token overhead across different methods. Here, the overhead means the ratio of the token cost of guard models compared with the token cost of backbone LLMs.}
    \label{tab:token_comparison}
\end{table*}

\end{document}